\documentclass{article}
\usepackage{spconf,amsmath,graphicx}
\usepackage{makecell}
\usepackage{multirow}
\usepackage{xcolor}
\usepackage{amsfonts}
\usepackage{booktabs}
\usepackage{verbatim}

\usepackage{eso-pic}
\AddToShipoutPicture{%
  \AtPageUpperLeft{%
    \raisebox{-1cm}[0pt][0pt]{\makebox[\paperwidth]{\centering
\scriptsize This work has been submitted to the IEEE for possible publication. Copyright may be transferred without notice, after which this version may no longer be accessible.
}}%
  }%
}

\title{HarmoniFuse: A Component-Selective and Prompt-Adaptive Framework for Multi-Task Speech Language Modeling}

\name{Yuke Si$^{1,2,\dagger}$, Runyan Yang$^{1,2,\dagger}$, Yingying Gao$^{1,2}$, Junlan Feng$^{1,2}$, Chao Deng$^{1,2}$, Shilei Zhang$^{1,2,*}$\\ \thanks{† Equal contribution}\thanks{* Corresponding author}}
\address{ $^1$Jiutian Artificial Intelligence Research Institute, China Mobile, Beijing, China\\ 
$^2$The State Key Laboratory of Multimedia Information Processing, Peking University, Beijing, China
}

\begin{document}

%
\maketitle
\begin{abstract}
Recent advances in large language models have facilitated the development of unified speech language models (SLMs) capable of supporting multiple speech tasks within a shared architecture. However, tasks such as automatic speech recognition (ASR) and speech emotion recognition (SER) rely on distinct types of information: ASR primarily depends on linguistic content, whereas SER requires the integration of both linguistic and paralinguistic cues. Existing multitask SLMs typically adopt naive parameter sharing or prompt-based conditioning without explicitly modeling the differences in information composition required by each task. Such designs risk task interference and performance degradation, especially under limited data conditions. To address these limitations, we propose \textbf{HarmoniFuse}, a component-selective and prompt-adaptive framework for multi-task speech language modeling. HarmoniFuse is designed to harmonize heterogeneous task demands by selecting and fusing task-relevant components of speech representations. Specifically, it integrates a gated speech encoder to extract task-specific acoustic features and a prompt-adaptive dynamic fusion module to aggregate transformer layers based on task characteristics. In addition, a batch-interleaved training strategy enables leveraging separate ASR and SER datasets without requiring joint annotation. Experimental results demonstrate that HarmoniFuse improves both ASR and SER performance, offering a scalable and robust solution for multitask speech understanding under realistic data constraints.
\end{abstract}

\section{Introduction}
\label{sec:intro}

Benefiting from the rapid evolution of large language models (LLM), large-scale speech understanding systems have increasingly adopted a unified multitask processing paradigm~\cite{ao2021speecht5, tang2023salmonn, deshmukh2023pengi,chu2023qwenaudio,chu2024qwen2audio,an2024funaudiollm}. These works demonstrate that shared encoder-decoder architectures can handle diverse speech tasks through multitask joint training, usually including automatic speech recognition (ASR), language identification, and speech emotion recognition (SER), etc. These unified multitask frameworks offer significant advantages by reducing deployment overhead, enhancing cross-task generalization, and enabling holistic speech understanding. 

Current large speech language models (SLMs) commonly adopt a brute-force multitask training strategy, leveraging hundreds of thousands of hours' annotated data to implicitly resolve task conflicts across diverse objectives~\cite{radford2022robustspeechrecognitionlargescale}. Some models further incorporate prompt-based decoding mechanisms to distinguish between tasks during inference~\cite{chu2023qwenaudio, an2024funaudiollm}, providing a lightweight form of task conditioning. However, these approaches typically lack explicit architectural mechanisms for modeling the divergent nature of tasks like ASR and SER. Moreover, the effectiveness of this data-driven strategy hinges on the availability of large-scale multi-label datasets, which are especially costly and scarce for affective tasks such as SER. This limits the scalability and applicability of current approaches in scenarios where the available data remains insufficient for large-scale model training.

In contrast, prior work on ASR-SER integration has largely been conducted in small-scale models, where researchers have proposed strategies to mitigate task interference. Classical approaches include cascaded pipelines that apply ASR followed by text-based SER~\cite{li2022fusing,li2024speech}, but such designs suffer from ASR error propagation and disconnect between acoustic and emotional features. More recently, some SER-related studies have introduced representation disentanglement~\cite{xi2022frontend,yuan2024disentanglement} and multi-stream attention mechanisms~\cite{yu2024speech, khan2025memocmt,zou2022speech} to separately model affective and linguistic cues. While these methods show promise in controlled settings, they are typically developed for specific tasks and lightweight model architectures, and have not yet been applied to large-scale SLMs with unified multitask capabilities.

To address the above challenges, we propose HarmoniFuse, a component-selective and prompt-adaptive framework for multi-task speech language modeling. The core insight is that effective multi-task learning for divergent tasks such as ASR and SER requires more than parameter sharing. It demands task-aware, dynamic control over how information is extracted, routed, and fused within the model. HarmoniFuse selectively emphasizes task-specific components based on task prompt, such as linguistic details for ASR and a combination of linguistic and paralinguistic cues for SER.

HarmoniFuse introduces three key innovations. First, a speech encoder selectively emphasizes task-relevant components from raw audio based on task characteristics.
Second, a prompt-adaptive layer fusion module dynamically aggregate transformer layers based on task prompts and input samples, allowing the decoder to flexibly shift attention between linguistic precision and paralinguistic expressiveness. Third, a batch-interleaved training strategy is introduced to overcome the scarcity of jointly annotated ASR-SER data, allowing the model to alternate between independent ASR and SER samples during training. This design leverages separate task-specific datasets while still promoting cross-task learning. Together, these innovations provide fine-grained control over task-aware information flow, effectively mitigating task interference, improving generalization across diverse tasks, and enabling robust multi-task speech understanding in data-constrained conditions.


\section{Method}\label{sec:method}
\begin{figure}[t]
\centering
\includegraphics[width=9cm]{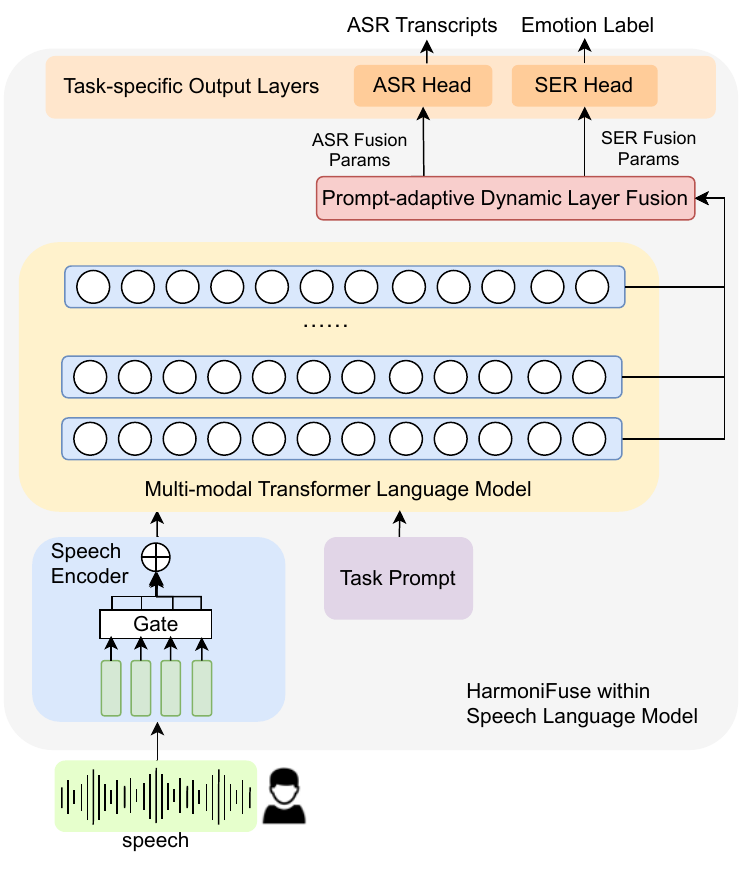}  
\caption{Overall architecture of proposed HarmoniFuse, featuring gated encoder-layer fusion and prompt-adaptive dynamic transformer layer fusion to harmonize linguistic and paralinguistic components across ASR and SER tasks.}
\label{fig1}
\end{figure}

As shown in Fig.~\ref{fig1}, HarmoniFuse consists of four components: (1) a speech encoder for extracting continuous acoustic representations from raw audio, (2) a multi-modal transformer language model for task-aware contextual processing, (3) a prompt-adaptive dynamic layer fusion module, and (4) task-specific output layers for downstream ASR or SER.

\subsection{Speech Encoder with Gated Layer Fusion}
The first stage of HarmoniFuse converts raw waveforms into high-level speech representations. We employ a pre-trained 24-layer WavLM model~\cite{chen2022wavlm} as the speech encoder due to its strong capability in capturing both phonetic content and prosodic cues across multiple layers. Given an input speech waveform, the speech encoder extracts a sequence of layer hidden states: $H_1, H_2, \ldots,H_{24} \in \mathbb{R}^{T_0 \times d_0}$, where $T_0$ is the representation sequence length and $d_0$ is the WavLM model hidden dimension.
To enable task-aware fusion across layers, we introduce a learnable weighting mechanism over all layers instead of using only the static layer:
\begin{equation}
H_{\rm fused}=\sum_{i=1}^{24} \frac{{\rm exp}\left(w_i\right)}{\sum _{j=1}^{24}{\rm exp}\left(w_j\right)} H_{i},
\end{equation}
where $w_i$ is the learnable weight parameter of the $i$-th layer, jointly optimized with other model parameters during end-to-end training. 
This layer fusion mechanism allows the encoder to dynamically emphasize different layer combinations based on task characteristics, thereby avoiding over-reliance on shallow or top-level features. It also enhances training efficiency and task specialization without requiring separate fine-tuning for each task.

\subsection{Multi-modal Transformer LM}
The second stage processes fused speech representations using a transformer-based language model. Inspired by recent multitask prompting strategies~\cite{yang2024polyspeech, an2024funaudiollm}, we design a structured input format comprising three main parts:
(1) speech representation $H_{\rm fused}$, (2) a task prompt token or token sequence (e.g., ``speech recognition'' for ASR or ``emotion recognition'' for SER), and (3) a ``BOS'' (beginning-of-sentence) token that signals the model to begin generation. These are concatenated and fed into a stack of $M$ Transformer encoder layers. 
Given the input sequence, the model produces each layer's contextual representations $R_{m}, \ m \in \left\{1, 2, \dots,  M\right\}$:
\begin{equation}
R_{m} = \left[\mathbf{r}_{m,1}, \mathbf{r}_{m,2}, \ldots, \mathbf{r}_{m,T}\right] \in \mathbb{R}^{T \times d}, 
\end{equation}
where $T$ is the sequence length and $d$ is the hidden dimension.


\subsection{Prompt-adaptive Dynamic Layer Fusion}
To enable task-specific adaptation, we introduce a dynamic layer fusion module on top of the SLM. For a given speech task $\tau$, at each output time step $t$, we use a set of weights to fuse the model's layer-wise representation:
\begin{equation}
 \alpha^\tau_t = \left[\alpha_{1,t}^\tau, \alpha_{2,t}^\tau, \dots, \alpha_{M,t}^\tau\right],   
\end{equation}
where each $\alpha_{m,t}^\tau \in \left[0,1\right]$ represents the importance of layer $m$. 
The final task-conditioned representation is computed as:
\begin{equation}
\mathbf{r}_t^\tau =\sum_{m=1}^{M}\alpha _{m,t}^{\tau}\mathbf{r}_{m,t}.
\end{equation}
Specifically, $\alpha_m^\tau$ for each layer is decomposed into two parts: 
\begin{equation}
\label{eq4}
\alpha_{m,t}^\tau = \beta \cdot \sigma\left(\lambda_m^\tau\right) + \left(1-\beta\right) \cdot \sigma\left({\rm FFN}\left(\mathbf{r}_{m,t}^\tau;\Theta_m\right)\right),
\end{equation}
where $\beta$ is a hyperpameter set to $0.5$, $\sigma\left(\cdot\right)$ denotes the sigmoid function, $\lambda_m^\tau$ is a learnable task- and layer-specific scalar parameter, and ${\rm FFN}\left(\cdot;\Theta_m\right)$ is a feed-forward network specific to layers with hidden dimention $d'=d/4$, taking hidden representation $\mathbf{r}_{m,t}^\tau$ as input. The first part of $\alpha_{m,t}^\tau$ encodes the preference directly related to the task, while the second part captures the relevance of speech representations as well as task prompts. 

\subsection{Task-specific Output Layers}
The final stage includes task-specific heads that project the fused transformer representation to output spaces for ASR or SER.

\subsubsection{ASR Head (Autoregressive Sequence Generation)}
For ASR, the model generates output tokens in an autoregressive manner. At each decoding step $t$, the head computes a probability distribution over the vocabulary using a linear projection followed by a softmax:
\begin{equation}
p_t\left(y_t\right) = {\rm Softmax}\left(W_{\rm ASR} \mathbf{r}_t^{\rm ASR} + b_{\rm ASR}\right)
\end{equation}
where $\mathbf{r}_t^{\rm ASR}$ is the output fused hidden state of the Transformer at step $t$, and $W_{\rm ASR}$, $b_{\rm ASR}$ are learnable parameters. The loss is computed using cross-entropy with teacher forcing during training.

\subsubsection{SER Head (Classification)} 
For SER, the SLM predict only the emotion label (i.e. $T=1$). So we use the fused hidden state $\mathbf{r}_{\rm 1}^{\rm SER}$ of the Transformer to predict the label. The head applies a fully connected layer:
\begin{equation}
p_{\rm SER} = {\rm Softmax}\left(W_{\rm SER} \mathbf{r}_{\rm 1}^{\rm SER} +b_{\rm SER}\right)
\end{equation}
where $W_{\rm SER}$, $b_{\rm SER}$ are task-specific learnable parameters. The loss is computed using categorical cross-entropy.

\section{Experiment setup}\label{sec:exp_setup}

\subsection{Datasets}\label{sec:data}
For the ASR task, we utilize the widely adopted 960 hours LibriSpeech dataset ~\cite{panayotov2015librispeech} as our primary training corpus. To enhance the robustness of speech recognition across variable speaking styles, we apply 3-fold speed perturbation at ratio 0.9, 1.0 and 1.1~\cite{ko2015audio}.

For the SER task, we adopt the IEMOCAP dataset~\cite{busso2008iemocap} as the primary supervised corpus. To scale up the data for large model training, we further incorporate unlabeled speech from the Emilia dataset~\cite{he2024emilia}, which was originally designed for speech synthesis and features rich emotional expressiveness. We follow the Emotion2Vec framework~\cite{ma2023emotion2vec} to automatically generate emotion pseudo-labels for Emilia. To ensure label compatibility with IEMOCAP, we perform emotion category alignment and retain four prototypical emotions: happy, sad, angry, and neutral~\cite{ma2023emotion2vec,chen2023exploring}. From the full Emilia corpus, we select approximately 1,800 hours of English utterances with a relatively balanced distribution across these four emotion classes. This pseudo-labeled subset of Emilia is then combined with IEMOCAP to form the training data. 
Instead of performing 5-fold cross-validation as commonly done for IEMOCAP in smaller-scale setups, we directly adopt the session 5 as a fixed held-out test set, in line with several existing works where high computational cost limits repeated training across folds~\cite{chen2023exploring}. The remainder of IEMOCAP data is used for training and validation, providing a rigorous and reproducible evaluation while avoiding unnecessary retraining.

\subsection{Model and Training Settings}\label{sec:expsetup}
The multi-modal transformer language model in HarmoniFuse consists of 12 transformer blocks, each with 12 attention heads, a hidden size of 768, and a feed-forward network size of 2,048. The total number of trainable parameters is approximately 160 million.
Training is performed on 4 NVIDIA A800 GPUs (80 GB each). One training batch on each GPU contains 400 seconds of speech, and we apply gradient accumulation over 4 steps to simulate a larger effective batch size.
Optimization is conducted using the Adam optimizer with an initial warm-up of 2,000 steps, followed by linear decay. We train the model for 20 epochs, and select the checkpoint with the lowest validation loss for final evaluation.

\section{Results and discussion}\label{sec:result}

We evaluate HarmoniFuse on two downstream tasks: SER and ASR. For SER, we report unweighted accuracy (UA) and weighted accuracy (WA), which measure performance by averaging over classes and by accounting for class distribution, respectively. For ASR, we report word error rate (WER) on both the ``clean'' and ``other'' subsets of the LibriSpeech test set. The results are presented in Tables~\ref{tab1} and~\ref{tab2}. 

\textbf{Single-task performance: gated layer aggregation enhances encoder representation learning.}
Table~\ref{tab1} shows the performance of different encoder layer selections in single-task settings. Prior studies have shown that different layers of pre-trained speech models like wav2vec 2.0 and HuBERT capture distinct types of information~\cite{de2024layer, pepino2021emotion}. They showed that lower and middle layers tend to preserve acoustic and paralinguistic features, while top layers encode more linguistic information. Our experiments corroborate these findings.

Specifically, we observe that using the 12th layer of the WavLM encoder yields the higher accuracy for SER (75.01\% on UA and 73.65\% on WA), likely because this layer best preserves emotional cues such as pitch, rhythm, and tone. In contrast, the 24th (final) layer achieves the lowest WER for ASR (4.5\% on clean, 6.0\% on other), consistent with its stronger focus on semantic content. To move beyond fixed-layer selection, we introduce a learnable gating mechanism to aggregate information across all encoder layers. This approach outperforms both single-layer baselines, improving SER to 75.31\% (UA) and 75.66\% (WA) and reducing ASR WER to 3.8\% (clean) and 5.1\% (other), showing that gated speech encoder is beneficial even in single-task scenarios.

\begin{table}[]
\caption{Experimental results on single-task models.}\label{tab1}
\renewcommand{\arraystretch}{1.2}
\resizebox{1\linewidth}{!}{%
\begin{tabular}{lcccc}
\hline
\multirow{2}{*}{Method} & \multicolumn{2}{c}{SER}              &ASR clean &ASR other    \\ \cline{2-5} 
                    & UA(\%)      &WA(\%)      & WER(\%)     & WER(\%)  \\ \hline
Encoder\_WavLM12        & 75.01     &73.65      & 8.0        & 13.9       \\
Encoder\_WavLM24        & 71.58     &71.64      & 4.5      & 6.0          \\
Encoder\_Gate           & 75.31     &75.66      & 3.8      & 5.1       \\ \hline       
\end{tabular}%
}
\end{table}

\textbf{Multi-task performance: handling conflicts between SER and ASR.}
Table~\ref{tab2} reports performance under multi-task learning with task prompts. It shows that the shared encoder alone cannot ensure effective multi-task learning, and the model exhibits trade-off results without the prompt-adaptive dynamic layer fusion. Specifically, the model using the 12th layer performs reasonably well on SER (75.51\% on UA and 74.94\% on WA) but poorly on ASR (6.1\% WER on clean and 12.1\% on other), while the 24th layer does the opposite. The gated speech encoder attempts to compromise, but still fails to reach the performance of the best single-task models, suggesting that static fusion cannot resolve task interference.

In contrast, HarmoniFuse introduces prompt-adaptive dynamic layer fusion, allowing the model to dynamically aggregate different layers depending on the task. This significantly improves both tasks: SER accuracy rises to 76.51\% (UA) and 76.31\% (WA), and ASR WER drops to 3.5\% (clean) and 5.1\% (other). Notably, both ASR and SER performances surpasses the single-task oracle baselines (WER 3.8\% on clean and 5.1\% on other for ASR, UA 75.31\% and WA 75.66\% for SER), confirming that our method effectively balances task-specific needs without requiring separate models.

\begin{table}[]
\caption{Experimental results on multi-task with task prompt.}\label{tab2}
\renewcommand{\arraystretch}{1.2}
\resizebox{1\linewidth}{!}{%
\begin{tabular}{lcccc}
\hline
\multirow{2}{*}{Method} & \multicolumn{2}{c}{SER}             & ASR clean   &ASR other    \\ \cline{2-5} 
                    & UA(\%)      &WA(\%)         & WER(\%)       & WER(\%)   \\ \hline
Encoder\_WavLM12    & 75.51  &74.94  & 6.1 & 12.1 \\ 
Encoder\_WavLM24    & 74.01  &73.97  & 3.8 & 5.6 \\ 
Encoder\_Gate       & 76.14  &75.42  & 3.9 & 5.8        \\

HarmoniFuse & \textbf{76.51} & \textbf{76.31}  & \textbf{3.5}     & \textbf{5.1}   \\   
\hline 
\end{tabular}%
}
\end{table}

\begin{figure}[]
\centering
\includegraphics[width=7.5cm]{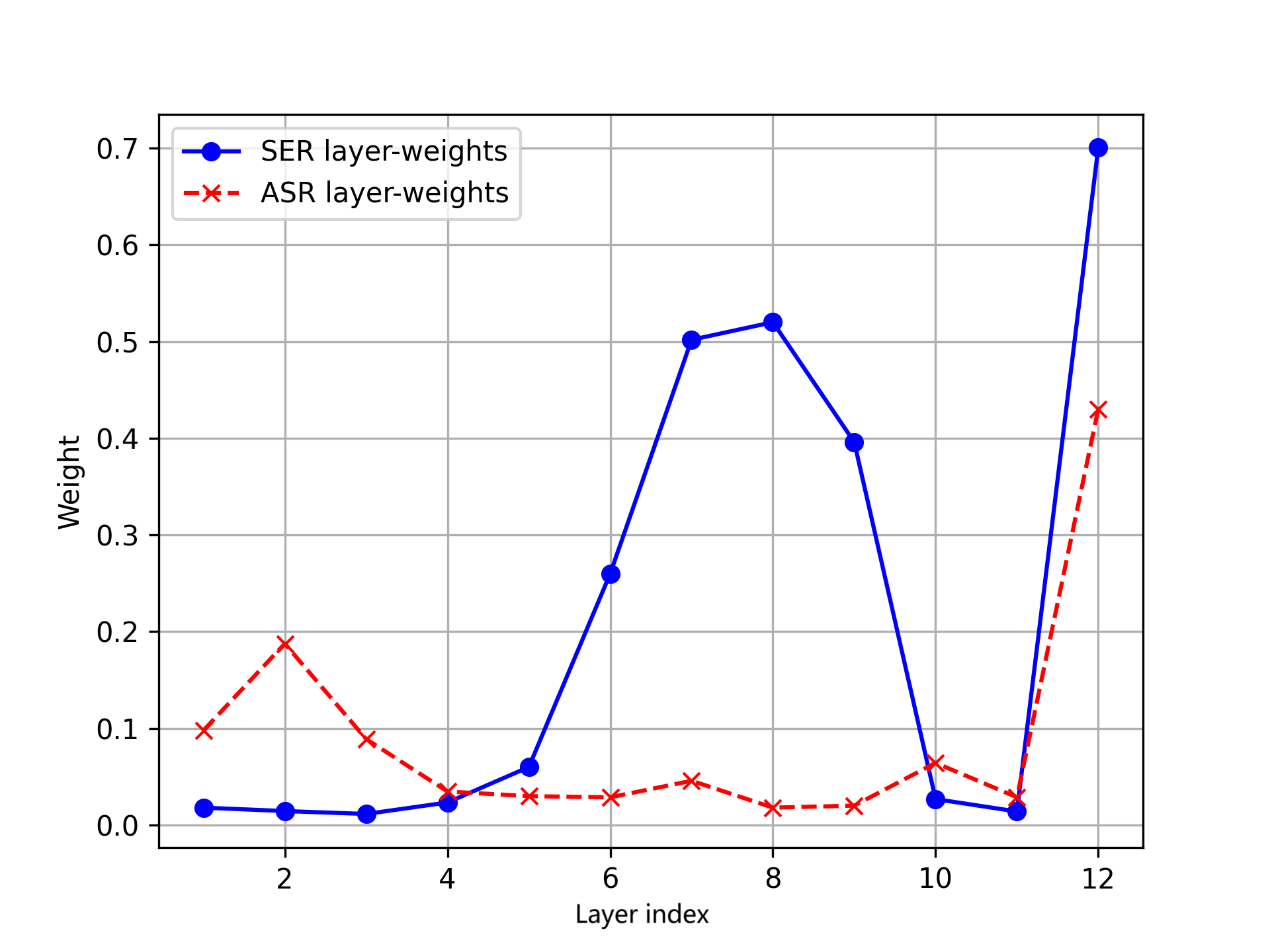} 
\caption{Comparison of layer-weights distribution between SER and ASR in speech language model.}
\label{fig2}
\end{figure}

To further analyze the model's behavior, we visualize the learned layer fusion weights $\sigma\left(\lambda_m^\tau\right)$ of Eq.~(\ref{eq4}) in Figure~\ref{fig2}. For SER, the model assigns higher importance to both the highest and mid-level layers (e.g., peaks at index 8 and 12), indicating a strong reliance on both prosodic and linguistic components. In contrast, the ASR task emphasizes deeper layers, more concentrating on refined linguistic component. It validates that HarmoniFuse learns task-discriminative representations by selecting the most relevant layers for each task. Rather than forcing a compromise, the model routes information according to the unique demands of each objective.

\section{Conclusion}\label{sec:conclusion}
In this work, we presented HarmoniFuse, a component-selective and prompt-adaptive framework designed to harmonize divergent objectives in multitask speech understanding, specifically focusing on ASR and SER tasks. The proposed framework incorporates three key innovations: a gated speech encoder, prompt-adaptive dynamic layer fusion, and batch-interleaved training strategy. These components jointly address longstanding challenges in multitask learning, including representation conflicts, limited joint annotations, and inter-task interference.
Experimental results demonstrate that it achieves strong performance across both tasks by dynamically coordinating linguistic and paralinguistic components, highlighting the benefits of task prompt-aware fusion in large speech language models.

\small
\bibliographystyle{IEEE}
\bibliography{mybib}

\end{document}